\documentclass[aps,pra,twocolumn,groupedaddress,nofootinbib]{revtex4-1}
\usepackage{bm}
\usepackage[dvipdfmx]{graphicx}
\usepackage[dvipdfmx]{color}
\usepackage{here}
\usepackage{amsmath,amsfonts,amssymb}
\usepackage{comment}
\usepackage{xcolor}
\usepackage{float}
\begin{document}
\title{
Band structures of generalized eigenvalue equation and conic section
}

\author{Takuma Isobe$^{1,*}$}
\author{Tsuneya Yoshida$^{2,3}$}
\author{Yasuhiro Hatsugai$^{1,4}$}
\affiliation{
$^1$Graduate School of Pure and Applied Sciences, University of Tsukuba, 
Tsukuba, Ibaraki 305-8571, Japan\\
$^2$Department of Physics, Kyoto University, Kyoto 606-8502, Japan,\\
$^3$Institute for Theoretical Physics, ETH Zurich, 8093 Zurich, Switzerland,\\
$^4$Department of Physics,
University of Tsukuba, Tsukuba, Ibaraki 305-8571, Japan
}
\renewcommand{\thefootnote}{}
\footnote{$^*$ Present adress: {\it Technology Development Headquaters, KONICA MINOLTA, Hachioji, Tokyo 192-8505, Japan.}}
\renewcommand{\thefootnote}{\arabic{footnote}}

\date{\today}

\begin{abstract}

Band structures of several metamaterials are described by generalized eigenvalue equations where complex bands emerge even if the involved matrices are Hermitian.
In this paper, we provide a geometrical understanding of the real-complex transition of the band structures. Specifically, 
our analysis, based on auxiliary eigenvalues, elucidates the correspondence between the real-complex transition of the generalized eigenvalue equations and Lifshitz transition in electron systems.
Furthermore, we elucidate that real (complex) bands of a photonic system correspond to the Fermi surfaces of type-II (type-I) Dirac cones in electron systems when the permittivity $\varepsilon$ and the permeability $\mu$ are independent of frequency.
In addition, our analysis elucidates that EPs are induced by the frequency dependence of the permittivity $\varepsilon$ and the permeability $\mu$ in our photonic system.
\end{abstract}

\maketitle

\section{Introduction}

The topology of band structures attaches growing interest~\cite{C.L.Kane_E.J.Mele_PRL.2005,C.L.Kane_E.J.Mele_PRL.2005_Z2,L.Fu_C.L.Kane_PRL.2007,M.Z.Hasan_C.L.Kane_RevModPhys.2010,X.L.Qi_S.C.Zhang_RevModPhys.2011,Y.Ando_JPSJ_2013,B.A.Bernevevig_T.L.Huglhes_S.C.Zhang_Science_2006,M.Knig_Science_2007,L.Fu_C.L.Kane_PRB.2007,L.Fu_C.L.Kane_PRB.2006,D.J.Thouless_PRB.1983,Schnyder_PRB.2008,A.Y.Kitaev_AIP_Conf_2009,S.Ryu_A.P.Schnyder_A.Furusaki_New.J.Phys_2010,X.L.Qi_T.L.Hughes_S.C.Zhang_PRB_2008,A.M.Essen_J.E.Moore_D.Vanderbilt_PRL_2009,S.Murakami_IOP_2007,W.Xiang_PRB_2011,Yang_PRB_2011,A.Birkov_PRL_2011,Xu_PRL_2011,Kurebayashi_JPSJ_2014,N.Armitage_RevModPhys_2018,Koshino_PRB_2016}. 
While it is originally developed in electron systems, topological bands in classical systems are extensively studied~\cite{Raghu_PhC_PRL(2008),Raghu_PhC_PRA(2008),MIT_PhChIns_PRL(2008),Lu_TopPhot_Nat(2014),Hu_TopPhot_PRL(2015),Takahashi_Optica(2017),Takahashi_JPSJ(2018),Ozawa_TopPhot_RMP19,OtaIwamoto_NatPhoto(2020),Moritake_NanoPh(2021),Kariyado_MechGraph_Nat(2015),Yang_TopAco_PRL(2015),Huber_TopMech_Nat(2016),Susstrunk_MechClass_PNAS(2016),Tomoda_AIP(2017),Takahashi_Mech_PRB(2019),Liu_TopPhon_AFM(2020),Lee_TopCir_Nat(2018),Yoshida_Difus_Nat(2021),Hu_ObsDifs_AM(2022),Knebel_GameTheor_PRL(2020),Yoshida_GameTheor_PRE(2021)}. Studies of topological bands in classical systems are accelerated by the recent development of non-Hermitian topology~\cite{Gong_class_PRX18,Kawabata_gapped_PRX19,Bergholtz_EP_RMP2021,Ashida_nHReview_AdvPhys2020} which exhibits unique topological bands such as exceptional points~\cite{Shen_EP_PRL2018,Zhen_ERing_nature(2015),Kozii_nH_arXiv(2017),Shen_NHTopBand_PRL(2018),Yoshida_NHhevferm_PRB(2018),Zyuzin_NHWyle_PRB(2018),Takata_pSSH_PRL(2018),Budich_SPERs_PRB(2019),Yoshida_SPERs_PRB(2019),Yoshida_SPERsMech_PRB(2019),Okugawa_SPERs_PRB(2019),Zhou_SPERs_Optica(2019),Mandal_HighEP_PRL(2021),Delplace_EP3_PRL(2021),IYH_PRB(2021),IYH_Nanoph(2023),Zhen_ERing_nature(2015),Takata_pSSH_PRL(2018),Budich_SPERs_PRB(2019),Yoshida_SPERs_PRB(2019),Yoshida_PTEP(2020),Yoshida_SPERsMech_PRB(2019),Okugawa_SPERs_PRB(2019),Zhou_SPERs_Optica(2019),Kawabata_gapless_PRL19}. These topological bands in metamaterials are based on a mathematical analogy of Schr\"{o}dinger equation (i.e., a standard eigenvalue equation).

Notably, some of the classical systems are beyond the standard eigenvalue equations and show their own developments. For instance, real-complex band structures of photonic systems can be understood in terms of generalized eigenvalue equations (GEVEs)~\cite{IYH_PRB(2021),IYH_Nanoph(2023),Yokomizo_GEVP_PRB2024}. In these systems, the indefinite property of the matrices induces complex band structures and exceptional points even when the involved matrices are both Hermitian.

Furthermore, the topological perspective is applied to nonlinear systems~\cite{Sone_TopoSync_PRR(2022),Sone_NLPsiCh_NatPhys2024,Sone_NLCaos_arXiv2024,IYH_NLEVP_PRL(2023), YIH_arxiv(2024)}. Prime examples of systems with nonlinearity of eigenvalues are photonic systems where the permittivity and permeability are frequency-dependent. Topological edge modes of these nonlinear systems are not straightforwardly understood from the bulk-edge correspondence of quantum systems~\cite{Hatsugai_BEC_PRL(1993),Hatsugai_BEC_PRB(1993)} described by standard eigenvalue equations. 
However, introducing auxiliary eigenvalues $\lambda$ elucidates the bulk-edge correspondence for the systems of nonlinear eigenvalue equations~\cite{IYH_NLEVP_PRL(2023)}; treating frequency (i.e. eigenvalues) as parameters, one can obtain physical bands as the section of $\lambda=0$ plane and auxiliary bands.

The above approach of auxiliary eigenvalues provides a geometrical understanding of band structures beyond standard eigenvalue equations. However, so far, applications of this approach have been limited to real bands.

In this paper, we explore complex band structures of systems described by a GEVE with Hermitian matrices.
Our analysis elucidates the correspondence between 
the real-complex transitions in systems of the GEVE and Lifshitz transition in electron systems, which provides a geometrical understanding of the complex bands. 
We further apply our approach to a photonic system. 
As a first step, we analyze the band structure by supposing that the permittivity $\varepsilon$ and the permeability $\mu$ are independent of the frequency $\omega$. 
Our analysis elucidates that real (complex) bands of the photonic systems correspond to Fermi surfaces of a type-II (type I) Dirac cone in electron systems~\cite{Kobayashi_DP-I_JPSJ(2007), Kawarabayashi_PRB(2011), Soluyanov_DPII_Nature(2015), Hatsugai_PRB(2015), Volovik_DPIII_JETP(2016)}. 
Our analysis taking into account the $\omega$ dependence elucidates that $\omega$ dependence of $\varepsilon$ and $\mu$ induces EPs.

The rest of this paper is organized as follows.
In Sec.~\ref{sec: toy model}, we analyze a toy model in terms of auxiliary eigenvalues, as well as a brief review of the GEVE.
In Sec.~\ref{sec: photonic}, we apply our approach to a photonic system.
A summary is provided in Sec.~\ref{sec: summary}. 
Appendix~\ref{sec: photo cont app} is devoted to a derivation of a GEVE for the photonic system.

\section{Conic sections and complex band structure of generalized eigenvalue equations}
\label{sec: toy model}

\begin{figure}[t]
   \includegraphics[width=9cm]{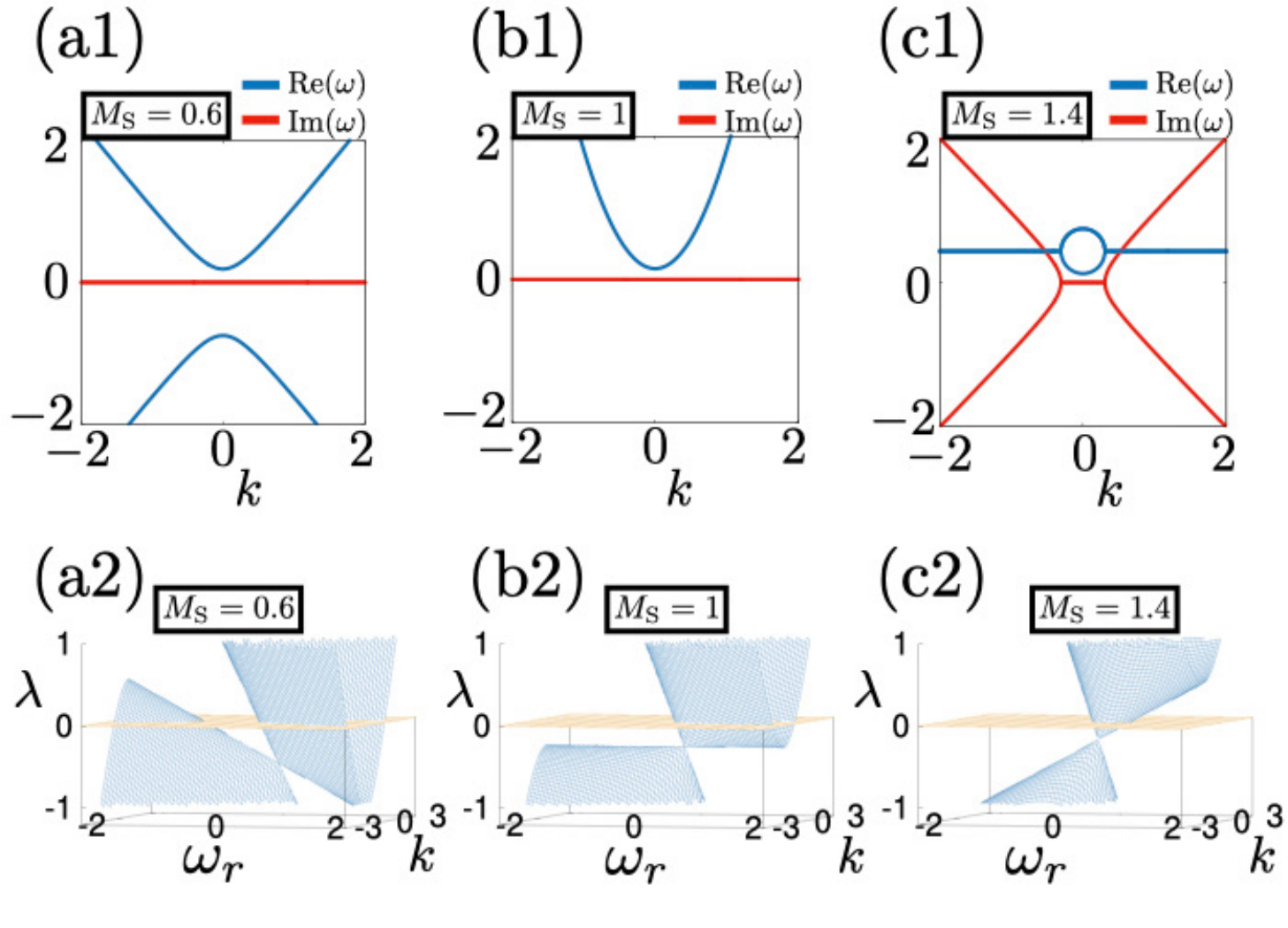}
 \caption{
Band structures of physical and auxiliary bands.
(a1)-(c1): Band structures of $\omega$ for each $k$ with $M_{\mathrm{S}}=0.6$, $M_{\mathrm{S}}=1$, and $M_{\mathrm{S}}=1.4$, respectively.
The real (imaginary) part of $\omega$ is plotted in blue (red).
(a2)-(c2): Band structures of $\lambda$ for each $\omega_{\mathrm{r}}$ and $k$ with $M_{\mathrm{S}}=0.6$, $M_{\mathrm{S}}=1$, and $M_{\mathrm{S}}=1.4$, respectively.
The yellow planes represent the plane of $\lambda=0$.
}
\label{fig: toy}
\end{figure}

Systems described by a GEVE may exhibit complex band structures and exceptional points due to the indefinite property of Hermitian matrices~\cite{IYH_PRB(2021),IYH_Nanoph(2023)}. 
Here, we provide their geometrical understanding by introducing auxiliary bands.
Our analysis elucidates the analogy between real-complex transitions of the band structure of a toy model described by the GEVE and Lifshitz transitions of electron systems.

Firstly, let us 
review complex bands induced by indefinite property of GEVE. We consider the system described by the following GEVE,
\begin{equation}
\label{eq: GEVP toy}
\begin{pmatrix}
M_{\mathrm{H}}&k\\
k&-M_{\mathrm{H}}
\end{pmatrix}\psi
=
\omega
\begin{pmatrix}
1+M_{\mathrm{S}}&0\\
0&1-M_{\mathrm{S}}
\end{pmatrix}
\psi.
\end{equation}
The eigenvalue $\omega$ is given by,
\begin{equation}
\label{eq: w toy}
\omega(k)=\frac{1}{1-M_{\mathrm{S}}}\left[-M_{\mathrm{S}}M_{\mathrm{H}}\pm\sqrt{M_{\mathrm{H}}^2+(1-M_{\mathrm{S}}^2)k^2}\right].
\end{equation}

When the involved matrices are indefinite, 
exceptional points and complex bands may emerge
even when the involved matrices are both Hermitian~\cite{indifiniteness_ftnt}. 
The band structures of this model are plotted in Fig.~\ref{fig: toy}(a1)-\ref{fig: toy}(c1) for $M_{\mathrm{H}}=0.3$. 
The real (imaginary) part of $\omega$ is plotted in blue (red).
When $M_{\mathrm{S}}<1$, eigenvalues $\omega$ become real [see Fig.~\ref{fig: toy}(a1)].
When $M_{\mathrm{S}}=1$, one of the bands diverges into infinity [see Fig.~\ref{fig: toy}(b1)].
Finally, when $M_{\mathrm{S}}>1$, $\omega$ becomes complex [see Fig.~\ref{fig: toy}(c1)].
In this case, exceptional points emerge at the band touching points of blue and red bands.

The above complex band structures are understood in terms of the auxiliary eigenvalues and conic sections which are analogous to various types of Dirac cones and the Fermi surfaces~\cite{Kobayashi_DP-I_JPSJ(2007), Kawarabayashi_PRB(2011), Soluyanov_DPII_Nature(2015), 
Hatsugai_PRB(2015), Volovik_DPIII_JETP(2016)}. 
We map the GEVE [Eq.~(\ref{eq: GEVP toy})]
to standard eigenvalue equation of auxiliary eigenvalues $\lambda$
\begin{equation}
\label{eq: P　psi=　lambda psi}
P(\omega,k)\psi=\lambda\psi,
\end{equation}
with 
\begin{equation}
P(\omega,k)=
\begin{pmatrix}
-\omega+M_{\mathrm{H}}-\omega M_{\mathrm{S}}&k\\
k&-\omega-M_{\mathrm{H}}+\omega M_{\mathrm{S}}&
\end{pmatrix}.
\end{equation}
We solve the above eigenvalue equation by regarding $\omega$ as a parameter. 
The physical band structure [see Fig.~\ref{fig: toy}] is obtained as the section with $\lambda=0$ plane.

Here, we discuss band structures of the auxiliary eigenvalues and their section with the $\lambda=0$ plane, which provides a geometrical understanding of band structures of the real part [$\omega_\mathrm{r}=\mathrm{Re}(\omega)$] and the 
exceptional point
induced by the indefinite property.

Diagonalizing matrix $P(\omega_{\mathrm{r}})$, we obtain
\begin{equation}
\lambda(\omega_{\mathrm{r}},k)=-\omega_{\mathrm{r}}\pm\sqrt{(M_{\mathrm{S}}\omega_{\mathrm{r}}-M_{\mathrm{H}})^2+k^2}.
\end{equation}
This model shows conical structures of the auxiliary bands [see Figs.~\ref{fig: toy}(a2)-\ref{fig: toy}(c2)].
For $M_{\mathrm{S}}<1$, the plane of $\lambda=0$ crosses both auxiliary bands $\lambda(\omega_{\mathrm{r}},k)$.
Thus, the physical band structure becomes Fig.~\ref{fig: toy}(a1).
Specifically, the conic section is described by
\begin{equation}
\frac{(\omega_{\mathrm{r}}+\alpha)^2}{A_1^2}-\frac{k^2}{B_1^2}=1,
\end{equation}
with
\begin{eqnarray}
&& A_1^2 = \frac{M_{\mathrm{H}}^2}{(1-M_{\mathrm{S}}^2)^2}, \quad
B_1^2 = 
\frac{M_{\mathrm{H}}^2}{|1-M_{\mathrm{S}}^2|}, 
\quad
\alpha =
\frac{M_{\mathrm{H}}M_{\mathrm{S}}}{1-M_{\mathrm{S}}^2}, \nonumber 
\end{eqnarray}
which leads hyperboloid of $\mathrm{Re}[\omega(k)]$ for $M_{\mathrm{S}}<1$.
Here, we can see that Fig.~\ref{fig: toy}(a1) corresponds to type-II Dirac cone~\cite{Soluyanov_DPII_Nature(2015)} by recognizing the auxiliary bands as energy bands of electrons and $\lambda=0$ plane as the Fermi energy.

For $M_{\mathrm{S}}=1$, the plane of $\lambda=0$ becomes parallel to the generatrix of the cone structure [see Fig.~\ref{fig: toy}(b2)].
Thus, only one band crosses the plane of $\lambda=0$ [see Fig.~\ref{fig: toy}(b1)].
Specifically, the conic section is described by
\begin{equation}
\label{eq: w k paraboid}
A_2\omega_{\mathrm{r}}+B_2k^2+C_2=0,
\end{equation}
with 
\begin{eqnarray}
&& A_2 = 2M_{\mathrm{H}},\quad 
B_2=-1,\quad 
C_2 = -M_{\mathrm{H}}^2. \nonumber 
\end{eqnarray}
Thus, the band structures of $\omega$ form paraboloid when $M_{\mathrm{S}}=1$.
Here, we can see that Fig.~\ref{fig: toy}(b2) corresponds to type-III Dirac cone~\cite{Volovik_DPIII_JETP(2016)} by recognizing the auxiliary bands as energy bands of electrons and $\lambda=0$ plane as the Fermi energy.

For $M_{\mathrm{S}}>1$, the plane of $\lambda=0$ cross only the upper band [see Fig.~\ref{fig: toy}(c2)].
Thus, the plane of $\lambda=0$ cut the cone structure in elliptic shape [see Fig.~\ref{fig: toy}(c1)].
Specifically, this ellipsoid is described by,
\begin{equation}
\label{eq: ellipsoid toy}
\frac{(\omega_{\mathrm{r}}+\alpha)^2}{A_3^2}+\frac{k^2}{B_3^2}=1,
\end{equation}
with 
\begin{eqnarray}
&& A_3^2 = \frac{M_{\mathrm{H}}^2}{(1-M_{\mathrm{S}}^2)^2}, \quad
B_3^2 = 
\frac{M_{\mathrm{H}}^2}{|1-M_{\mathrm{S}}^2|}, 
\quad
\alpha =
\frac{M_{\mathrm{H}}M_{\mathrm{S}}}{1-M_{\mathrm{S}}^2}, \nonumber 
\end{eqnarray}
Equation~(\ref{eq: ellipsoid toy}) elucidates that band touching of the real part with vanishing the imaginary part. 
This band touching is nothing but exceptional points induced by the indefinite property.

Here, we can see that Fig.~\ref{fig: toy}(c2) corresponds to type-I Dirac cone~\cite{Kobayashi_DP-I_JPSJ(2007), Kawarabayashi_PRB(2011), Hatsugai_PRB(2015)} by recognizing the auxiliary bands as energy bands of electrons and $\lambda=0$ plane as the Fermi energy.

In the above, we have analyzed complex band structures of the system described by GEVE [see Eq.~(\ref{eq: GEVP toy})] in terms of auxiliary bands and their conic section.

\section{Application to a photonic system}
\label{sec: photonic}
\subsection{Conics in the Maxwell equations}
\begin{figure}[t]
 \includegraphics[width=9cm]{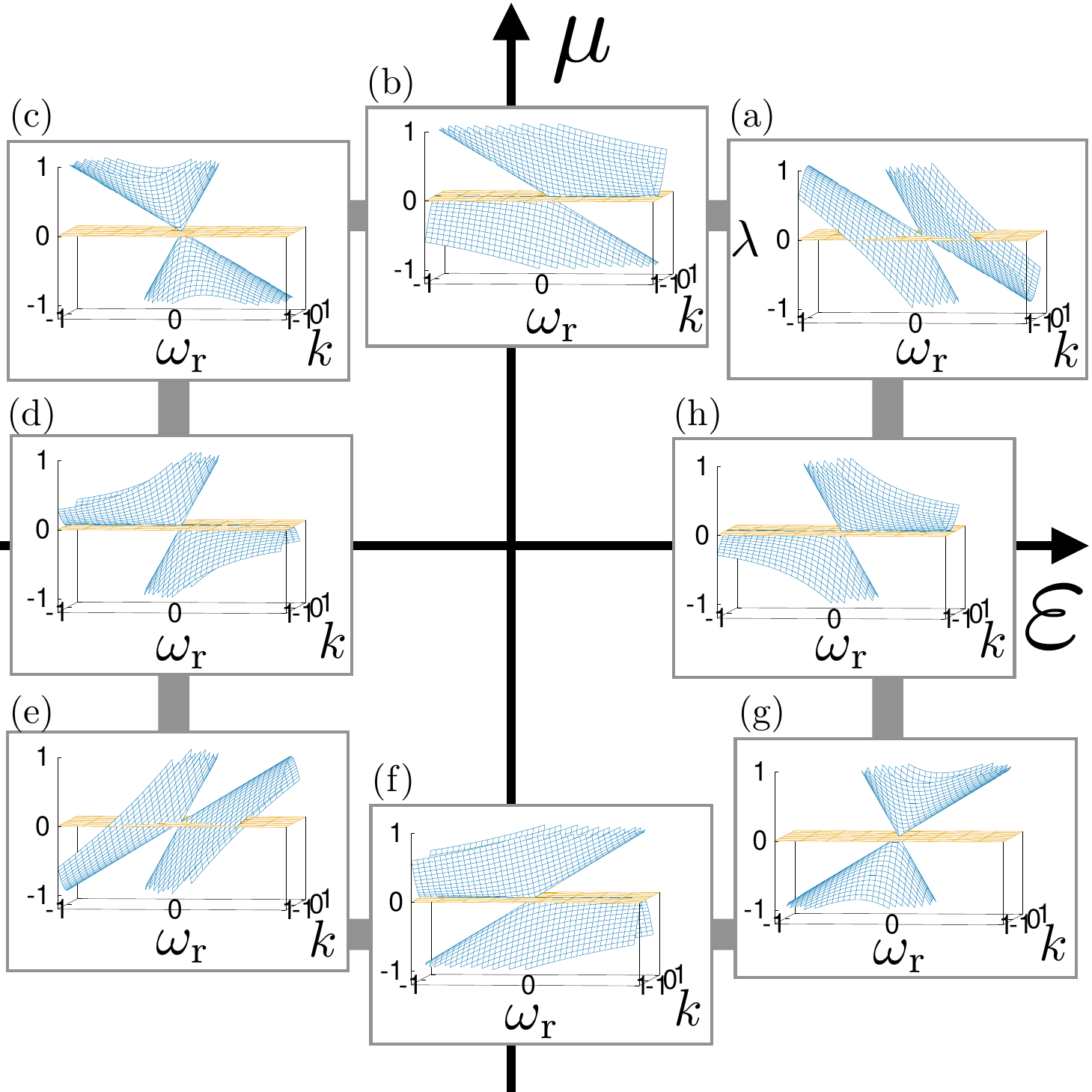}
 \caption{
Band structures of $\lambda$ for each $\omega_{\mathrm{r}}$ and $k$.
 The yellow plane represents the plane of $\lambda=0$.
Panels (a), (b), (c), (d), (e), (f), (g), and (h) are obtained for $(\varepsilon,\mu)=(3,1)$, $(0,1)$, $(-3,1)$, $(-3,0)$, $(-3,-1)$, $(0,-1)$, and $(3,-1)$, respectively.
 }
\label{fig: photo cycle}
\end{figure}

Here, let us apply the above discussion to photonic systems.
We discuss the one-dimensional continuum media, described by the Maxwell equation,
\begin{equation}
\label{eq: photo cont}
\begin{pmatrix}
0&k\\
k&0
\end{pmatrix}\psi
=
\omega
\begin{pmatrix}
\varepsilon&0\\
0&\mu
\end{pmatrix}
\psi,
\end{equation}
where $\varepsilon$ and $\mu$ represent the permittivity and the permeability.
The eigenvector $\psi$ corresponds to the electromagnetic field.
The derivation of Eq.~\eqref{eq: photo cont} is provided in Appendix~\ref{sec: photo cont app}.

Here, we suppose that $\varepsilon$ and $\mu$ are independent of $\omega$. The $\omega$ dependence  is taken into account in Sec.~\ref{sec: cont photo w dep}. In the case of this system, diagonal components of the left-hand side matrix become zero.

Here, band structures are given by the following dispersion relation,
\begin{equation}
\label{eq: w-k const ep mu}
\omega(k)=\pm \sqrt{\frac{k^2}{\varepsilon\mu}}.
\end{equation}
Band structures in this model are plotted in Fig.~\ref{fig: photo cycle}.
Equation~(\ref{eq: w-k const ep mu})
indicates that band structures $\omega(k)$ are given by real when both the permittivity and the permeability are positive or negative.
On the other hand, band structures becomes pure-imaginary, when either the permittivity or the permeability is negative.
This result is consistent with the fact that the electromagnetic fields cannot penetrate into the single negative media such as metals~\cite{Maier_plasmonics_Springer(2007)}.

Next, let us discuss the band structures of $\lambda$ [see Eq.~(\ref{eq: P　psi=　lambda psi})] by analyzing
\begin{equation}
\begin{pmatrix}
-\omega\varepsilon&k\\
k&-\omega\mu
\end{pmatrix}
\psi=\lambda\psi.
\end{equation}
Here, the auxiliary eigenvalues are given by,
\begin{equation}
\lambda(\omega,k)=-\frac{\omega}{2}(\varepsilon+\mu)\pm\sqrt{\left(\frac{\omega}{2}\right)^2(\varepsilon-\mu)^2+k^2}.
\end{equation}
Band structures of $\lambda$ are plotted in Fig.~\ref{fig: photo cycle} for several values of $\varepsilon$ and $\mu$.
In this model, cone structures of $\lambda$ emerge in $\omega$-$k$ space.

When $\varepsilon$ and $\mu$ are located in the first or third quadrant, 
the auxiliary bands cross the plane specified by $\lambda=0$ [see Figs.~\ref{fig: photo cycle}(a)~and~\ref{fig: photo cycle}(e)].  
Thus, the eigenvalues $\omega(k)$ become real.
In this case, the band structure of $\lambda$ corresponds to the type-II Dirac cone where auxiliary bands and the $\lambda=0$ plane correspond to the energy band of electrons and the Fermi energy, respectively. Here, $\omega(k)$ corresponds to the Fermi surface. When $\varepsilon$ and $\mu$ locate in the second or fourth quadrant, the auxiliary bands touch the plane specified by $\lambda=0$ only at point $(\omega_{\mathrm{r}},k)=(0,0)$ [see Figs.~\ref{fig: photo cycle}(c)~and~\ref{fig: photo cycle}(g)].
Thus, the eigenvalues $\omega(k)$ become pure imaginary.
In this case, the band structure of $\lambda$ corresponds to the type-I Dirac cone.
When either of $\varepsilon$ or $\mu$ is zero,
generatrix of two cones touches $\lambda=0$ [see 
Figs.~\ref{fig: photo cycle}(b), 
\ref{fig: photo cycle}(d),
\ref{fig: photo cycle}(f),
and \ref{fig: photo cycle}(h)], 
which analogous to the type-III Dirac cone.

The above analysis based on axially eigenvalue $\lambda$ elucidates that 
real-complex transition of band structures $\omega(k)$ corresponds to changes in types of Dirac cones.
For our model, the upper and the lower bands touch at $\lambda=0$ due to vanishing diagonal elements of the matrix on the left-hand side of Eq.~(\ref{eq: photo cont}). Thus, our photonic model does not exhibit an EP (see Fig.~\ref{fig: photo cycle}) in contrast to the toy model [see Eq.~(\ref{eq: GEVP toy}) and Fig.~\ref{fig: toy}].
We note, however, that the $\omega$ dependence of $\varepsilon$ and $\mu$ induces EPs as discussed in Sec.~\ref{sec: cont photo w dep}.

\begin{figure}[]
   \includegraphics[width=9cm]{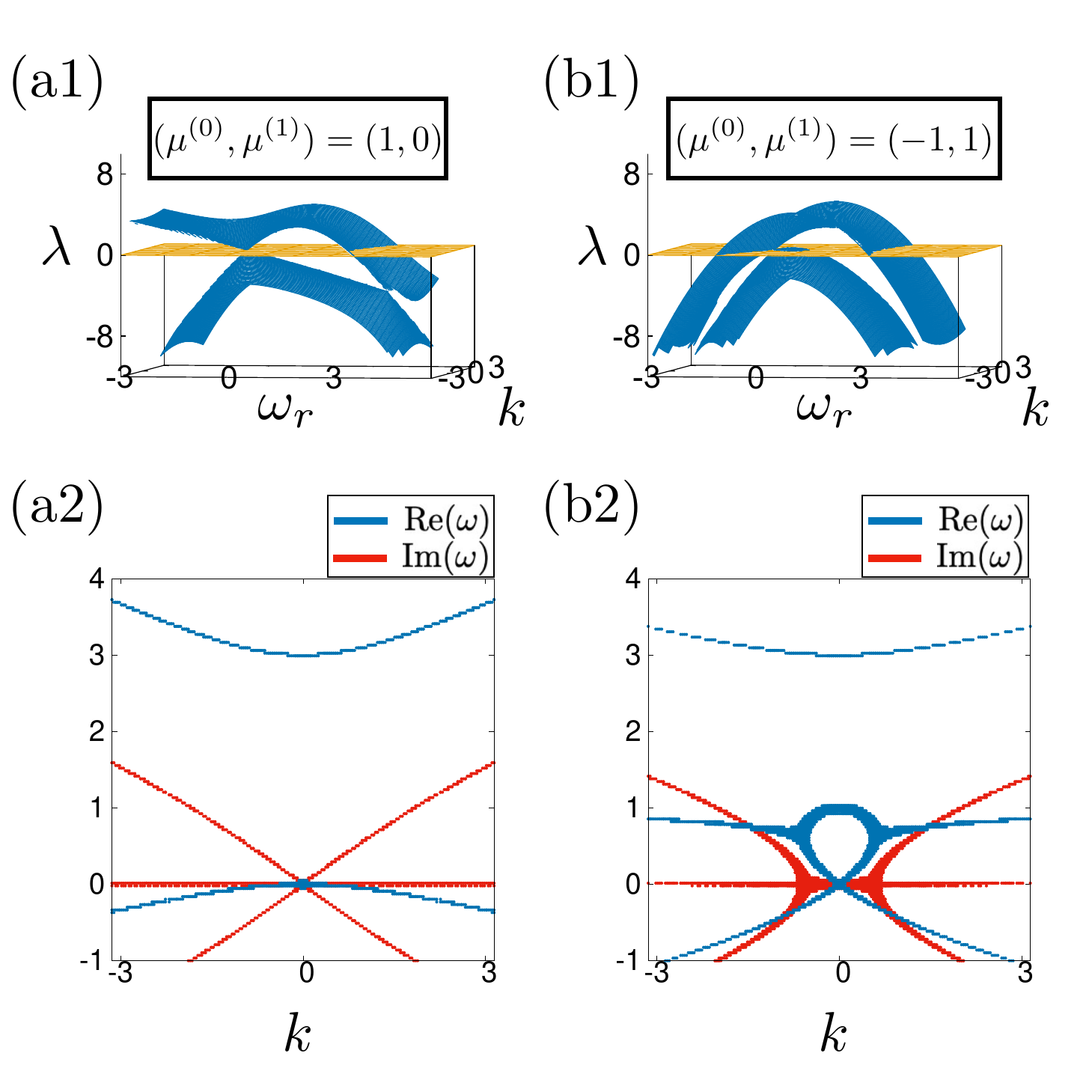}
 \caption{
Analysis of the effect of the $\omega$-dependence of $\varepsilon$ and $\mu$.
(a1)-(b1): Plot of the auxiliary band structure of $\lambda$.
The yellow surface represent $\lambda=0$.
(a2)-(b2): Plot of  the physical band structures of $\omega$.
The blue (red) lines represent the real (imaginary) part of $\omega$.
The yellow surface represents $\lambda=0$.
Panels~(a1)~and~(a2) [(b1)~and~(b2)] are obtained for $(\varepsilon^{(0)},\varepsilon^{(1)}, \mu^{(0)}, \mu^{(1)},\omega_a)
=(-3,1,1,0,0 )$ [$(-3,1,-1,1,0)$].
}
\label{fig: photo cont}
\end{figure}

\subsection{
Effect of the $\omega$-dependence of 
$\varepsilon$ and 
$\mu$
}
\label{sec: cont photo w dep}

We analyze effects of the frequency dependence of the permittivity 
$\varepsilon$ and the permeability $\mu$, which elucidates the emergence of EPs.
We consider the $\omega$-dependence
up to the first order of $\omega$
\begin{align}
\varepsilon(\omega)\approx\varepsilon^{(0)}+\varepsilon^{(1)}(\omega-\omega_a),\\
\mu(\omega)\approx\mu^{(0)}+\mu^{(1)}(\omega-\omega_a).
\end{align}
Using these equations, Eq.~(\ref{eq: photo cont}) becomes,
\begin{equation}
\begin{pmatrix}
0&k\\
k&0
\end{pmatrix}\psi
=
\omega
\begin{pmatrix}
\tilde{\varepsilon}^{(0)}&0\\
0&\tilde{\mu}^{(0)}
\end{pmatrix}
\psi
+\omega^2
\begin{pmatrix}
\varepsilon^{(1)}&0\\
0&\mu^{(1)}
\end{pmatrix}
\psi,
\end{equation}
where $\tilde{\varepsilon}^{(0)}=\varepsilon^{(0)}-\varepsilon^{(1)}\omega_a$ and $\tilde{\mu}^{(0)}=\mu^{(0)}-\mu^{(1)}\omega_a$.
In this case, matrix $P(\omega,k)$ is written as
\begin{equation}
P(\omega,k)=
\begin{pmatrix}
-\omega\tilde{\varepsilon}^{(0)}-\omega^2\varepsilon^{(1)}&k\\
k&-\omega\tilde{\mu}^{(0)}-\omega^2\mu^{(1)}\\
\end{pmatrix}.
\end{equation}

Analyzing the auxiliary band structure, we observe the emergence of EPs.
The band structures of $\lambda$ are plotted in Figs.~\ref{fig: photo cont}(a1)~and~\ref{fig: photo cont}(b1).
These data indicate that the auxiliary bands bend due to the $\omega$-dependence. This band bending of $\lambda$ leads to eigenvalues $\omega(k=0)$ taking real values [see Figs.~\ref{fig: photo cont}(a2)~and~\ref{fig: photo cont}(b2)], which is a qualitative difference from the results in the previous section [see Figs.~\ref{fig: photo cycle}(c)~and~\ref{fig: photo cycle}(d)].
In particular, for $(\mu^(0),\mu^(1))=(-1,1)$, the band bending of $\lambda$ results in the emergence of EPs [see Fig.~\ref{fig: photo cont}(b2)].

The above results indicate that the $\omega$ dependence of $\varepsilon$ and $\mu$ results in the bending of the auxiliary bands, leading to the emergence of EPs.
We stress that the analogy between real-complex transitions and Lifshitz transitions holds when
the permittivity and the permeability depend on $\omega$.

\section{Summary}
\label{sec: summary}
In this article, we have studied the real-complex transition of the band structures on the GEVEs with Hermitian matrices in terms of
auxiliary eigenvalues $\lambda$.
Our approach provides a geometrical understanding of the real-complex transition of the band structures, where the physical bands are obtained as the sections of auxiliary bands and $\lambda=0$ plane.

We have also applied our approach to the photonic system. Our analysis elucidates the analogy between the real-complex transition of the band structures in the photonic system and the Lifshitz transition of Dirac cones in electron systems; real (complex) bands of photonic systems correspond to the Fermi surface of type-II (I) Dirac cone.
In addition, we elucidate that the $\omega$ dependence of the permittivity and the permeability induces the emergence of EPs in our photonic system.


\section{ACKNOWLEDGEMENT}
This work is supported by JST-CREST Grant No.~JPMJCR19T1, JST-SPRING Grant No.~JPMJSP2124, and JSPS KAKENHI Grant No.~JP21K13850, JP23K25788, and JP23KK0247.
This work is also supported by JSPS Bilateral Program No.~JSBP120249925.
T.Y is grateful for the support from the ETH Pauli Center for Theoretical Studies and the Grant from Yamada Science Foundation.

\bibliography{main}

\begin{thebibliography}{10}
\bibitem{indifiniteness_ftnt}
Consider a GEVE $H\psi=\omega S\psi$ with Hermitian matrices $H$ and $S$ whose
  determinants are finite. Although the above GEVE is rewritten as a standard
  eigenvalue problem with a non-Hermitian matrix $S^{-1}H$, eigenvalues
  $\omega$ remain real when matrices $H$ and $S$ are definite.\\ For positive
  definite matrix $S$, a GEVE is reduced to a standard eigenvalue problem with
  Hermitian matrix $ \tilde{H} \Psi =\omega \Psi $. with
  $\tilde{H}=S^{-1/2}HS^{-1/2}$ and $\Psi=S^{1/2}\psi$. Here $S^{-1/2}$ is
  Hermitian matrix defined as $S^{-1/2}=U \mathrm{diag}(s_1,s_2,\ldots)
  U^\dagger$ with an unitary matrix diagonalizing $S$ and the eigenvalues $s_i$
  ($i=1,2,\ldots$). Thus, eigenvalues $\omega$ are real when $S$ is positive
  definite. A similar argument is available for negative definite $S$. In
  addition, replacing $\omega$ to $1/\omega$ leads us to real eigenvalues for
  definite matrix $H$. Therefore, complex eigenvalues $\omega$ require
  indefiniteness of both Hermitian matrices $H$ and $S$.
\end{thebibliography}
\bibliographystyle{h-physrev5}

\appendix

\section{
Derivation of Eq.~(\ref{eq: photo cont})
}
\label{sec: photo cont app}
In the continuum media, Maxwell equations can be transformed to the Dirac-like form.
In general, Maxwell equations are described by the GEVE composed of $6\times6$ matrices $H_{6\times 6}\psi=\omega S_{6\times 6\psi}$,
\begin{equation}
\label{eq:6x6maxwell}
\begin{pmatrix}
0&-\bm{k}\times\\
\bm{k}\times&0
\end{pmatrix}
\begin{pmatrix}
\bm{E}\\
\bm{H}
\end{pmatrix}
=\omega
\begin{pmatrix}
\varepsilon&\alpha\\
\alpha^{\dagger}&\mu
\end{pmatrix}
\begin{pmatrix}
\bm{E}\\
\bm{H}
\end{pmatrix}.
\end{equation}
Since this equation does not include Gauss's low, the solution include the longitudinal modes.

Here, let us consider reducing the matrix size of Eq.~(\ref{eq:6x6maxwell}).
At first, we employ the basis of the transverse magnetic (TM) modes and the transverse electric (TE) modes.
The basis can be transformed by using the following matrix $U_{6\times 6}$,
\begin{equation}
U_{6\times 6}
\bm{F}=
\left(
\begin{array}{ccc|ccc}
1&0&0&0&0&0\\
0&1&0&0&0&0\\
0&0&0&0&0&1\\ \hline
0&0&0&1&0&0\\
0&0&0&0&1&0\\
0&0&1&0&0&0
\end{array}\right)
\left(
\begin{array}{c}
E_x\\
E_y\\
E_z\\ \hline
H_x\\
H_y\\
H_z
\end{array}\right)=
\left(
\begin{array}{c}
E_x\\
E_y\\
H_z\\ \hline
H_x\\
H_y\\
E_z
\end{array}
\right)
\end{equation}

Therefore, the GEVE is transformed as follows,
\begin{equation}
\label{eq:TETM-Maxwell}
\left[U_{6\times6}H(\bm{k})U_{6\times 6}^{-1}\right]\left[U_{6\times 6}\bm{F}\right]=\omega \left[U_{6\times6}SU_{6\times 6}^{-1}\right]\left[U_{6\times 6}\bm{F}\right]
\end{equation}
with
\begin{equation}
U_{6\times6}H(\bm{k})U_{6\times 6}^{-1}=
\left(
\begin{array}{c|c}
\begin{matrix}
0&0&-k_y\\
0&0&k_x\\
-k_y&k_x&0
\end{matrix}
&
\begin{matrix}
0&k_z&0\\
-k_z&0&0\\
0&0&0 
\end{matrix} 
\\ \hline
\begin{matrix}
0&-k_z&0\\
k_z&0&0\\
0&0&0
\end{matrix}
&
\begin{matrix}
0&0&k_y\\
0&0&-k_x\\
k_y&-k_x&0
\end{matrix}
\end{array}\right),
\end{equation}
\begin{equation}
U_{6\times6}S(\bm{k})U_{6\times 6}^{-1}=
\left(
\begin{array}{c|c}
\begin{matrix}
\varepsilon_{11}&\varepsilon_{12}&\alpha_{13}\\
\varepsilon_{21}&\varepsilon_{22}&\alpha_{23}\\
\beta_{31}&\beta_{32}&\mu_{33}
\end{matrix}
&
\begin{matrix}
\alpha_{11}&\alpha_{12}&\varepsilon_{13}\\
\alpha_{21}&\alpha_{22}&\varepsilon_{23}\\ 
\mu_{31}&\mu_{32}&\beta_{33}
\end{matrix}
\\ \hline
\begin{matrix}
\beta_{11}&\beta_{12}&\mu_{13}\\
\beta_{21}&\beta_{22}&\mu_{23}\\
\varepsilon_{31}&\varepsilon_{32}&\alpha_{33}
\end{matrix}
&
\begin{matrix}
\mu_{11}&\mu_{12}&\beta_{13}\\
\mu_{21}&\mu_{22}&\beta_{23}\\
\alpha_{31}&\alpha_{32}&\varepsilon_{33}
\end{matrix}
\end{array}\right), 
\end{equation}
Here, the left-hand side matrix can be block diagonalized when the system is two-dimensional.
Thus, when the off-diagonal block of the right-hand side matrix in Eq.~(\ref{eq:TETM-Maxwell}) is zero, the Maxwell equation can be decomposed
as follows
TE modes:
\begin{align}
\label{eq:3x3Maxwell}
\begin{pmatrix}
0&0&-k_y\\
0&0&k_x\\
-k_y&k_x&0
\end{pmatrix}
\begin{pmatrix}
E_x\\
E_y\\
H_z
\end{pmatrix}
&=\omega
\begin{pmatrix}
\varepsilon_{11}&\varepsilon_{12}&\alpha_{13}\\
\varepsilon_{21}&\varepsilon_{22}&\alpha_{23}\\
\beta_{31}&\beta_{32}&\mu_{33}
\end{pmatrix}
\begin{pmatrix}
E_x\\
E_y\\
H_z
\end{pmatrix}, 
\end{align}
TM modes:
\begin{align}
\begin{pmatrix}
0&0&k_y\\
0&0&-k_x\\
k_y&-k_x&0
\end{pmatrix}
\begin{pmatrix}
H_x\\
H_y\\
E_z
\end{pmatrix}
&=\omega
\begin{pmatrix}
\mu_{11}&\mu_{12}&\beta_{13}\\
\mu_{21}&\mu_{22}&\beta_{23}\\
\alpha_{31}&\alpha_{32}&\varepsilon_{33}
\end{pmatrix}
\begin{pmatrix}
H_x\\
H_y\\
E_z
\end{pmatrix}. 
\end{align}
These equations generally possess three eigenvectors. 
While two of them satisfy Gauss's law, one of them does not satisfy the law.
Projecting out the eigenspace that does not satisfy Gauss's law, the size of the matrices in the above equations can be further reduced when the system is isotropic.
Here, the eigenvector of the longitudinal mode is given by, $\bm{v}_0=(k_x,k_y,0)^T/\sqrt{k_x+ik_y}$.
In order to project out the eigenspace of the longitudinal mode, we employ two-vectors perpendicular to $v_0$,
\begin{equation}
\bm{v_1}=
\frac{1}{\sqrt{k_x+ik_y}}
\begin{pmatrix}
0\\
-k_y\\
k_x
\end{pmatrix}\ \ \mathrm{and} \ \
\bm{v_2}=
\begin{pmatrix}
0\\
0\\
1
\end{pmatrix}.
\end{equation}
Here, by using the basis $V=(\bm{v}_0,\bm{v}_1,\bm{v}_2)$, Eqs.~(\ref{eq:3x3Maxwell}) are transformed as $(V^{\dagger}H_{3\times 3}V)(V^{\dagger}\psi)=\omega (V^{\dagger}S_{3\times 3}V)(V^{\dagger}\psi)$.
Therefore, Eq.~(\ref{eq:3x3Maxwell}) becomes,
\begin{align}\label{Eq:2x2Maxwell}
\begin{pmatrix}
0&k_x-ik_y\\
k_x+ik_y&0
\end{pmatrix}
\bm{\psi}_{\mathrm{TE}}
&=\omega
\begin{pmatrix}
\varepsilon&0\\
0&\mu
\end{pmatrix}
\bm{\psi}_{\mathrm{TE}},\\
\notag \\
\begin{pmatrix}
0&k_x-ik_y\\
k_x+ik_y&0
\end{pmatrix}
\bm{\psi}_{\mathrm{TM}}
&=\omega
\begin{pmatrix}
\mu&0\\
0&\varepsilon
\end{pmatrix}
\bm{\psi}_{\mathrm{TM}}.
\end{align}

\end{document}